\documentclass[preprint2]{aastex}

\newcommand{\ARAA}{ARA\&A}

\def\kms{\mbox{km~s$^{-1}$}}

\def\kpch{\mbox{$h^{-1}$ kpc}}

\def\LCDM{\mbox{$\Lambda$CDM}}

\def\mpch{\mbox{$h^{-1}$ Mpc}}
\def\Mpch{\mbox{$h^{-1}$ Mpc}}
\def\Mvir{\mbox{M$_{\rm vir}$}}
\def\M2vir{\mbox{M$_{\rm 2vir}$}}
\def\Msun{\mbox{M$_\odot$}}
\def\Msunh{\mbox{$h^{-1}M_\odot$}}

\def\Rvir{\mbox{R$_{\rm vir}$}}

\begin{document}
\shortauthors{PRADA ET AL.}
\shorttitle{How far do they go?}
\title{How far do they go? The outer structure of dark matter halos}

\author{Francisco Prada$^1$, Anatoly A. Klypin$^2$, Eduardo Simonneau$^3$,
Juan Betancort-Rijo$^4$, Santiago Patiri$^4$, Stefan Gottl\"ober$^5$ and 
Miguel A. Sanchez-Conde$^1$} 

\affil{$^1$Instituto de Astrof\'\i sica  de 
Andaluc\'\i a (CSIC), Camino Bajo de Huetor, 50, E-18008 Granada, Spain}

\affil{$^2$Astronomy Department, New Mexico State University,
MSC 4500, P.O.Box 30001, Las Cruces, NM, 880003-8001, USA}

\affil{$^3$Institut d'Astrophysique de Paris, CNRS, 75014 Paris, France}

\affil{$^4$Instituto de Astrof\'\i sica de Canarias, V\'{i}a L\'{a}ctea 
s/n, E-38200 La Laguna, Tenerife, Spain}

\affil{$^5$Astrophysikalisches Institut Postdam, An der Sternwarte 16, 14482 Potsdam, Germany}




\begin{abstract}
We study the density profiles of collapsed galaxy-size dark matter
halos with masses $10^{11}-5\cdot 10^{12}\Msun$ focusing mostly on the
halo outer regions from the formal virial radius $\Rvir$ up to
5-7$\Rvir$.  We find that isolated halos in this mass range extend
well beyond $\Rvir$ exhibiting all properties of virialized objects up
to 2--3$\Rvir$: relatively smooth density profiles and no systematic
infall velocities. The dark matter halos in this mass range do not
grow as one naively may expect through a steady accretion of
satellites, i.e., on average there is no mass infall.  This is
strikingly different from more massive halos, which have large infall
velocities outside of the virial radius. We provide accurate fit for
the density profile of these galaxy-size halos.  For a wide range
$(0.01-2)\Rvir$ of radii the halo density profiles are fit with the
approximation $\rho=\rho_s \exp\left({-2n[x^{1/n}-1]}\right)
+\langle \rho_m \rangle$, where $x\equiv r/r_s$, $\langle \rho_m \rangle$ is the mean matter density
of the Universe, and the index $n$ is in the range $n=6-7.5$. These
profiles do not show a sudden change of behavior beyond the virial
radius.  For larger radii we combine the statistics of the initial
fluctuations with the spherical collapse model to obtain predictions for
the mean and most probable density profiles for halos of several
masses. The model give excellent results beyond 2-3 formal virial
radii.
\end{abstract}



\section{Introduction}
\label{sec:intro}

More than twenty years of extensive work on cosmological N-body
numerical simulations have provided numerous detailed predictions for
the structure of dark matter halos in the hierarchical clustering scenario.
\citet[][hereafter NFW]{NFW}
preceded by the pioneering efforts of
\citet{Quinn86,Frenk88,DC91,Warren92,Crone94}, suggested a simple
fitting formula to describe the spherically averaged density profile
of isolated dark matter halos in virial equilibrium. Since then
numerous simulations were done for many relaxed halos of different
masses and in different cosmologies. The NFW analytical density
profile
\begin{equation}
    \rho(r) =\frac{\rho_s}{x(1+x)},\quad x\equiv r/r_s  \label{eq:NFW}
\end{equation}
has two parameters: the characteristic density $\rho_{\rm s}$ and the radius
$r_{\rm s}$.  Instead of these parameters, one can use the virial mass of
the halo, $\Mvir$, and the concentration, $C\equiv\Rvir/r_{\rm s}$.
Here the mass $\Mvir$ and the corresponding radius $\Rvir$ are defined as
the mass and the
radius within which the spherically averaged overdensity is equal to
some specific value. For the standard cosmological model with the
cosmological constant $\LCDM$ and parameters 
$\Omega_{0}=0.3$, $\Omega_{\Lambda} = 0.7$, and $h = 0.7$ we have 
$\Mvir=4\pi(340\langle \rho_m \rangle)R^3_{\rm vir}/3$, 
where $\langle \rho_m \rangle$ is the average matter density in the Universe.  For a halo with 
this profile,
$\rho \propto r^{-1}$ as $r\rightarrow0$ and smoothly fall off as
$\rho \propto r^{-3}$ at the virial radius. The concentration
parameter weakly depends on the virial mass with a significant scatter
comparable to the systematic change in $C$ over three decades in
$\Mvir$ \citep{Bullock01,Eke01}.

Later simulations paid most of attention to the inner slope of the
profiles.  Some results favored a steeper profile than NFW density cusp with
$\rho \propto r^{-1.5}$ \citep{FM97,M98,JS00,G00}. More recent
simulations of halos with millions of particles within $\Rvir$ seem to
indicate that there is a scatter in the inner slope of the density
profiles across a wide range of masses -- from dwarfs to clusters. The
inner slope varies between these two shapes: the NFW with an asyntotic
slope of one and the steeper \citet{M98} with slope 1.5
\citep[see][]{Klypin01,Reed03,Navarro04,Diemand,WBO04,Tasitsiomi04,FKM04}.

Different approximations for density profiles were suggested and
tested in the literature. Just as some other groups, we find that the
3D S\'ersic three-parameter approximation gives extremely good fits
for dark matter halos \citep{Navarro04,Merritt2005}. We slightly
modify this approximation by adding the average matter density of the
Universe $\langle \rho_m \rangle$. This term can be neglected, if one fits the
density inside the virial radius. Yet, at larger distances, it gives
an important contribution. The approximation can be written  as
\begin{equation}
    \rho(r) = \rho_s \exp\left({-2n[x^{1/n}-1]}\right) +\langle \rho_m \rangle,\quad x\equiv r/r_s.  
\label{eq:Sersic}
\end{equation}
where  $n$ is the S\'ersic index.
  
In addition to all the numerical simulations, a significant effort has
been made to compare the predictions of the $\LCDM$ model with the
observations. This is the case of the most recent set of high quality
observations of large samples of rotation curves of galaxies or the
strong gravitational lensing studies which place an important upper
limit on the amount of dark matter in galaxies and clusters in the
inner few to tens of kiloparsec (within $r_{\rm s}$) where the need of
a cuspy density profile is still subject of an exciting debate
\citep[e.g.,][and references
therein]{FP94,M94,dB03,S03,Rhee03,Keeton98, Keeton01,Bro04}.

 On the theoretical side, however, the origin of the shape of the dark
matter halo density profile remains poorly understood. It is generally
accepted that the dark matter halos are assembled by hierarchical
clustering as the result of halo merging and continuous
accretion. This merging scenario has motivated an interest in the analysis
of the mass accretion history of the halos in conjunction with their
structural properties \citep[e.g.,][]{Wechsler}. The systematic study
of the NFW density fits to many simulated halos shows that their mass
accretion history is closely correlated with the concentration
parameter $C$ and, therefore, with the mass inside the scale radius
$r_{\rm s}$ \citep{Wechsler,Zhao,Tasitsiomi04}. These results suggest
that the formation process of the dark matter halos can be generally
understood by an early phase of fast mass accretion and a late phase
of slow accretion of mass. In this scenario, the inner dense regions
of the halos are build up early during the fast phase of mass
accretion when the halo mass increases with time much faster than the
expansion rate of the Universe. At later epochs, during the phase of
slow mass accretion, the outer regions of the halo are built, while
its inner regions stay almost intact \citep[see][]{Zhao}.

 Despite to all this effort dedicated to the understanding of the central
dense regions of the dark matter halos, very little attention has been
devoted to the study of their outskirts, i.e. the regions beyond the formal
virial radius. The outer parts of the halos and therefore their
density profiles exhibit in these regions large fluctuations which can
be understood as the result of infalling dark matter (including
infalling smaller halos or substructure) or due to major mergers. In
both cases the infalling material has not reached the equilibrium with
the rest of the halo \citep[see][]{FM01}. On the contrary, a
considerable observational effort is being made to measure the mass
distribution around galaxies and clusters at large distances using
weak gravitational lensing \citep[e.g.,][and references
therein]{Smith01,GS02,Kneib03,Hoekstra,Sheldon}. In these cases the distances go
well beyond the virial radius ranging from few hundred kpc to several Mpc.
Individual field galaxies or clusters produce a small distortion
of the background galaxies that allows us to measure the surface mass
density profile of the dark matter 
\citep[e.g.,][]{Mellier,BS01}. It is customary in the weak lensing
analysis to specify a dark matter halo density profile to model
the projected mass profile measured with this technique.  The NFW
analytical formula is often adopted and extrapolated at large
distances, beyond $\Rvir$ with $\rho \propto r^{-3}$. This density model
may not be accurate enough.

The motion of satellite galaxies as a test for dark 
matter distribution at
large radii \citep[e.g.,][]{ZaritskyWhite,Zaritsky97,
Prada2003,Brainerd04,DEEP} is another observational method, which
requires detailed theoretical predictions for outer density profiles
and infall velocities. 
 
In fact, we really do not know how far the halos extend.
Indeed, the NFW fitting formula was proposed and extensively tested to
describe dark matter halos within $\Rvir$.  This is why the NFW density
fits are always done within the virial radius or even well below the
virial radius ($<0.5\Rvir$), where the halos are expected to be
virialized in order to avoid such non-equilibrium fluctuations. In
this context, it is also surprising that the prediction of the
spherical collapse model for the mean profile, which can be reasonably
expected to give good predictions at sufficiently large distances
($>2\Rvir$), have not been used. In fact, the predictions of this
models have only recently been worked out \citep[see][]{Barkana}, but
they have not been tested against numerical simulations.

Our goal  is to carry out a detailed study of the density profiles
in and around collapsed objects.  
%
The paper is organized as follows. In Section~\ref{sec:num}, we give
the details of the numerical simulations and the fits of
the density profile of distinct galaxy-size halos.
In Section~\ref{sec:infall}, we have studied the shape of the density
and infall velocity profiles of isolated halos
up to $2-3\Rvir$ for different masses. We show in 
Section~\ref{sec:sph} that for large radii the mean density profile around 
dark matter halos is in excellent agreement with the predictions we have 
obtained via the spherical collapse model, which are somewhat different from 
those found by \citet{Barkana}. Finally, in Section~\ref{sec:fin} 
discussions and conclusions are given.

Throughout this paper the formal virial radius is the radius within which the mean 
matter density  is equal to 340 times the average mean matter 
density $\rho_m$ of the Universe at $z=0$.

\section{Numerical simulations and density fits}
\label{sec:num}

The simulations used in this paper are done using the Adaptive
Refinement Tree (ART) code \citep[ART,][]{Kravtsov97}. The simulations were done 
for the standard $\LCDM$ cosmological model with 
$\Omega_{0}=0.3$, $\Omega_{\Lambda} = 0.7$, and $h = 0.7$.
 We study  halos selected from four different 
simulations, which parameters are listed in 
Table~\ref{tab:tab1}.

The simulations cover a wide range of scales and have different mass
and force resolutions. The simulation Box20 has the highest
resolution, but it has only two galaxy-size halos.
We use them as examples for the structure of halos simulated with very
high resolution. In most of the cases we limit the analysis to well
resolved halos: those should have more than $\approx$20,000 particles
inside virial radius. The simulation Box120 has the larger volume,
but its mass resolution allows us to use only halos with masses larger
than $10^{13}\Msunh$. Most of the analysis of galaxy-size halos is
done using simulations Box80S and Box80G. In the case of the
simulation Box80G the whole $80\Mpch$ volume was resolved with
equal-mass particles. There were about 180,000 halos in the
simulation.  The simulation Box80S was done using particles with
different masses. Only a small fraction -- a $10\Mpch$ radius region
-- of the box was resolved with small-mass particles. The high resolution
region has an average density about equal to the mean density of the
Universe. It was chosen in such a way that the halo mass function in
the region is representative for the typical region of this size. For
example, the region does not have a massive cluster. The most massive
halo in the region has mass $2.6\times 10^{13}\Msunh$. There are 5
halos with mass larger than $10^{13}\Msunh$.  Altogether, there are
about 60,000 halos in this simulation.

The density profile of each halo, which we study, is fit by the approximation
given in eq.(~\ref{eq:Sersic}). Each fit provides the
concentration $C$ and the S\'ersic index $n$. We often average density profiles
of halos for some
range of $\Mvir$. When doing so, we scale the radii to units of the virial
radius of each halo and then average the densities. The averaged density profile
is then fit again.  In some cases, before we do the fitting, we also
split the halo population of a given mass range into 3-4 sub-samples
with a narrow range of concentrations. The parameters of the density fits
together with some other properties of the halos are given in
Table~\ref{tab:tab2}.

The halos in our catalogs come from different environments. Some of them
are inside the virial radii of larger halos; some have strong interactions
with smaller, but still massive neighbors. We call a halo ``distinct''
if it does not belong to a larger halo. Most of the time we are
interested in isolated halos: distinct halos, which do not have large
companions. We search for halos around the given halo. If within the
distance $d\times\Rvir$~ the largest companion is smaller than $\Mvir/m$,
then the halo is called isolated. When doing the pair-wise comparisons of
the halos, we use the largest virial radius of the two halos, but we
use $\Mvir$ of the given halo for the test of the masses.  Different
isolation criteria are used. We typically use the $d=2$ and $m=5$ combination
(no massive companion within 2$\Rvir$). For Milky-Way size halos with
$\Mvir \approx 10^{12}\Msunh$ this condition typically gives 50\%-60\%
of all distinct halos of this mass.

\begin{table*}[tb]
\tablenum{1}
\label{tab:tab1}
\caption{Parameters of Simulations}
\begin{center}
\small
\begin{tabular}{lccccl}
\tableline\tableline\\
{Name} & 
{Box} & 
{Mass resolution} &
{Force resolution} &
{Number of particles}  &
{Number of steps} 
\\
 &
\multicolumn{1}{c}{(\mpch)} & 
\multicolumn{1}{c}{(\Msunh)}&
\multicolumn{1}{c}{($h^{-1}$ kpc)} &
\multicolumn{1}{c}{($10^6$)} 
\\
\tableline
\\
Box20 & 20 & $6.1\times 10^{5}$ &0.15 & 9.0 & $5\times 10^5$  \\
Box80S& 80 & $4.9\times 10^{6}$ &0.15 & 160 & $1\times 10^6$         \\
Box80G& 80 & $3.2\times 10^{8}$ &1.2  & 134 & $1.25\times 10^5$ \\
Box120& 120 & $1.1\times 10^{9}$ &1.8  & 134 & $1.25\times 10^5$ \\
\\             
\tableline
\end{tabular}
\end{center}
\end{table*}

\begin{table*}[tb]
\tablenum{2}
\label{tab:tab2}
\caption{Parameters of halos}
\begin{center}
\small
\begin{tabular}{lllllll}
\tableline\tableline\\
\multicolumn{1}{c}{Model Name} & 
\multicolumn{1}{c}{Virial Mass}   &
\multicolumn{1}{c}{Number of halos} &
\multicolumn{1}{c}{Number of particles} &
\multicolumn{1}{c}{Concentration} &
\multicolumn{1}{c}{ $n$} & 
\multicolumn{1}{c}{Figure} 
\\
&
\multicolumn{1}{c}{(\Msunh)} & &
\multicolumn{1}{c}{inside $R_{\rm vir}$}
\\
\tableline
\\
Box20 & $1.4\times 10^{12}$      & 1  & $2.2\times 10^6$ & 14.3 & 5.9 & \ref{fig:fig1} \\
      & $2.8\times 10^{11}$      & 1  & $4.6\times 10^5$ & 16.7 & 7.5 & \ref{fig:fig1} \\
Box80S&$(1\pm 0.3)\times 10^{11}$& 162 & $(2\pm 0.7)\times 10^4$ & 18  & 7.0 & \ref{fig:fig3}\\
      &$(4\pm 2)\times 10^{11}$  & 79 & $(8\pm 4)\times 10^4$ & 14.0  & 6.5 &\ref{fig:fig3},\ref{fig:figNFW} \\
Box80G&$(6\pm 2)\times 10^{12}$ & 192& $(1.8\pm 0.6)\times 10^4$ & 11.7 & 6.3 & \ref{fig:fig3} \\
\\             
\tableline
\end{tabular}
\end{center}
\end{table*}

\section{Halo profiles and infall velocities}
\label{sec:infall}

\begin{figure}[tb!]
\plotone{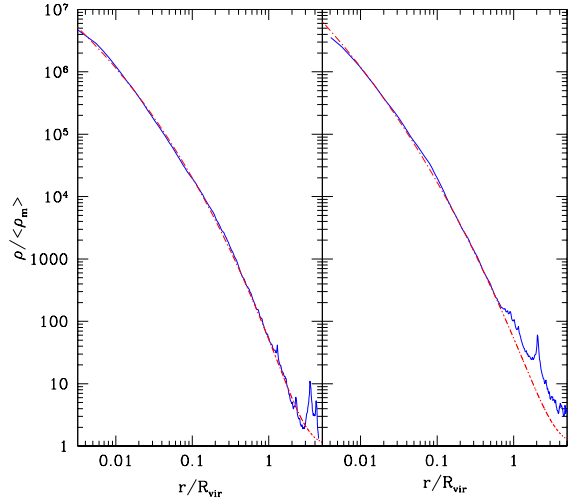}
\caption{Dark matter density profiles of two dark matter halos (full
curves) in the simulation Box20. The halos have virial masses of
$1.4\times 10^{12}\Msunh$ (left panel) and
$2.6\times 10^{11}\Msunh$ (right panel). The larger halo
has a neighbour at 3.5 $\Rvir$ which is the halo on the right panel. This
smaller halo is responsible for the spike at large radii in the density
profile. In turn, the halo on the right panel has its own smaller neigbour at
2$\Rvir$ observed as a spike and an extended bump in the density
profile.  The dashed curves show the 3D Sersic profiles. The halo density
profiles extend well beyond the formal virial radius with the S\'ersic
profile providing remarkably good fits.}
\label{fig:fig1}
\end{figure}

Figure~\ref{fig:fig1} gives examples of density profiles of two halos
with virial masses $1.4\times 10^{12}\Msunh$ (left panel) and
$2.6\times 10^{11}\Msunh$ (right panel) in the simulation Box20. 
The halos have the virial radii of $230\kpch$ and $130\kpch$ respectively. 
The halos were done
with very high resolution, which allows us to track the density
profile below $0.01\Rvir$. 
The larger halo is isolated with its nearest
companion being at $3.5\Rvir$. 
The density profile of the halo has some
spikes due to substructure, but otherwise it clearly extends up to
$3\Rvir$ where we see large fluctuations due to its companion. The
smaller halo on the right panel has a neighbor at $2\Rvir$. So, it is
not isolated. Eq.(~\ref{eq:Sersic}) gives very good approximations for
both halos. 

\begin{figure}[tb!]
\plotone{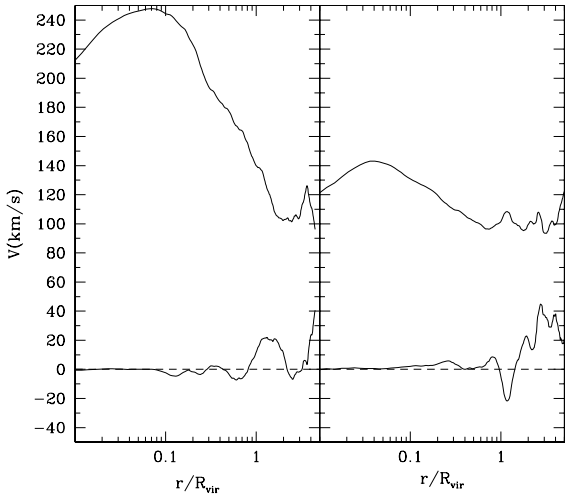}
\caption{The top curves on each panel show the 3D rms
velocities at different radii for the halos shown in
Figure~\ref{fig:fig1}. The bottom curves show the average
radial velocities. The radial velocities show that for isolated halos
of Milky-Way size and smaller halos there is no infall velocities outside
of formal virial radius.}
\label{fig:fig2}
\end{figure}

Figure~\ref{fig:fig2} gives more information on the structure of these
two halos. The 3D rms velocities are what one would naively expect for
``normal'' halos.  The rms velocity first increases when we go from
the center and reaches a maximum at some distance. The radius of the maximum rms
velocity is smaller for the halo on the right panel. This is because
it has larger concentration. At larger radii the rms velocity first
declines relatively smoothly. It has some fluctuations due to
substructure. At radii larger than $\Rvir$ the decline stops.  The
average radial velocity is more interesting and to some degree is
surprising. Nothing unusual inside $\Rvir$: it is practically zero with
tiny ($\approx 5\kms$) variations due to substructure. This is a clear
sign of a virialized object. At larger distances the fluctuations in
the velocity increase, but not dramatically if we compare those
fluctuations with the rms velocities. The smaller halo on the right
has a narrow dip ($V_{\rm rad}\approx -20\kms$) at $1.2\Rvir$ apparently
due to a satellite, which is moving into the halo. The surprising
result is what we {\it do not} find, i.e., there is no infall on the two
halos. This may seem like a fluke. Indeed, halos must grow. Their mass
must increase with time. In order for the mass to increase, there
should be on average negative infall velocities just outside the
virial radius. Yet, we do not find those. We will later see that these
two halos are not flukes, but are typical examples for halos of this mass
range. What is important at this stage is that the infall velocities
outside of $\Rvir$ are very small. They are significantly smaller than the rms
velocities. These small radial velocities indicate that halos may
extend to radii significantly larger than their formal virial radii.

\begin{figure}[tb!]
\epsscale{1.12}
\plotone{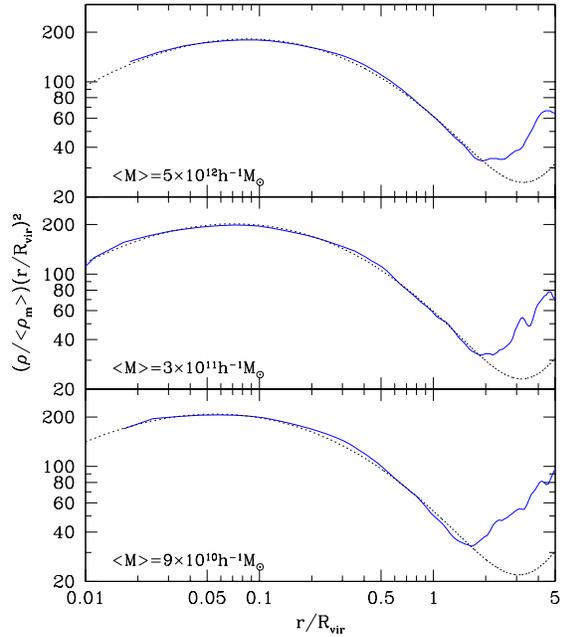}
\caption{Average density profiles for halos with different virial
masses. The 3D S\'ersic profile provides very good fit with few percent errors
within $2\Rvir$. Even at $3\Rvir$ the error is less than 20-30
percent. The density profiles are well above the average density of
the Universe throughout all the radii.}
\label{fig:fig3}
\end{figure}

The average density profiles of isolated halos of different masses are
shown in Figure~\ref{fig:fig3}. The isolation criteria in this case is
no massive satellite inside $2\Rvir$ ($d=2, m=5$; see Sec~\ref{sec:num}). The
smooth density profiles, which are extremely accurately fit by
eq.(~\ref{eq:Sersic}), extend from the smallest resolved radius all
the way to 2\Rvir. Parameters of the density fits are given in
Table~\ref{tab:tab2}. Note that in order to reduce the range of
variations along the y-axis, we plot density multiplied by
$(r/\Rvir)^2$. The horizontal parts of the curves in this plots correspond to
density declining as $\rho\propto r^{-2}$. The density profiles are
well above the average density of the Universe throughout all the
radii. Even at $5\Rvir$ the average density profile is still 4-5 times
larger than the mean density.  The upturn at large radii tells us that
the density declines less steep than $r^{-2}$.

\begin{figure}[tb!]
\epsscale{1.0}
\plotone{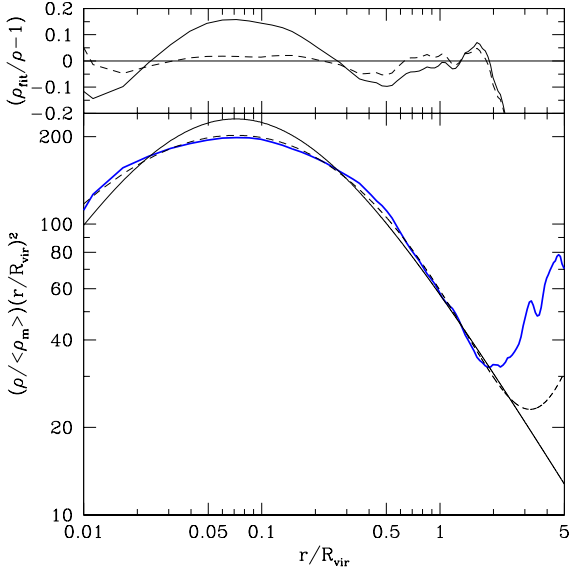}
\caption{Comparison of the NFW and the 3D S\'ersic fits.  The thick
full curve in the bottom panel shows the average density for halos
with mass $<M>=3\times 10^{11}\Msunh$. The 3D S\'ersic fit is the
dashed curve. The NFW fit is presented by the thin full curve. The top
panel shows the errors of the fits. The full curve is for the NFW
approximation, and the dashed curve is for the 3D S\'ersic fit. In
outer regions $r=(0.3-2)\Rvir$ both fits have practically the same
accuracy. Both fits start to fail at larger distances. Overall, the 3D
S\'ersic approximation provides remarkable accurate fit.}
\label{fig:figNFW}
\end{figure}

The NFW approximation provides less accurate fits as shown in
Figure~\ref{fig:figNFW}. One may attribute the success of the 3D
S\'ersic approximation to the fact that it has three free parameters,
while the NFW has only two. This is correct only to some degree.  The
problem with the NFW is that it has a slightly wrong shape at radii
around $r_s$: its curvature is a bit too large. In
Figure~\ref{fig:figNFW} this is manifested by an extended hump close
to the maximum of the curves at $r\approx (0.05-0.2)\Rvir$. One can
shift the NFW slightly to the right and down to make the fit more
accurate for most of the body of the halo ($r>0.05\Rvir$). In this
case the NFW fit goes below the halo density at small radii
$r<0.05\Rvir$, which sometimes was interpreted as if the central slope
is steeper than -1.  Overall, the 3D S\'ersic approximation provides
remarkable accurate fit. For $r=(0.01-2)\Rvir$ the errors are smaller
than 5\%.

\begin{figure}[tb!]
\epsscale{1.12}
\plotone{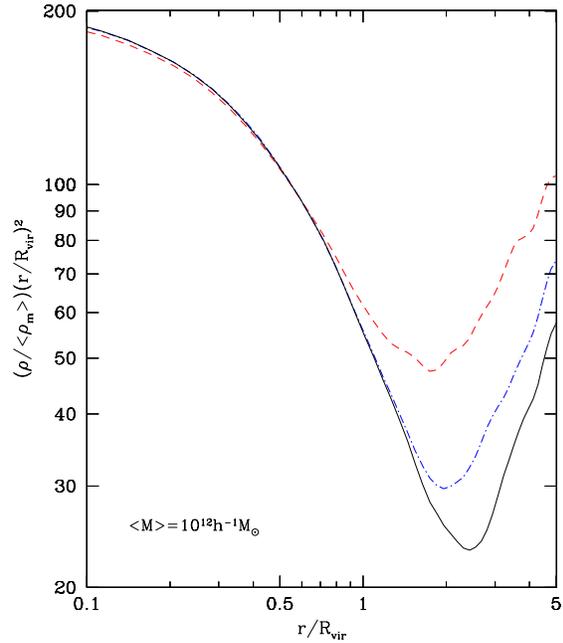}
\caption{Effects of different isolation criteria on the density
profiles.  The dashed curve shows the average density profile of
distinct halos with mass $M\approx 10^{12}\Msunh$ in Box80G.  The
dot-dashed and full curves are for isolated halos, which do not have
massive companions inside $2\Rvir$ and $3\Rvir$ respectively.  More strict isolation
conditions result in lower density outside of the virial radius.  At
all shown radii the densities are well above the average density of
the Universe.}
\label{fig:isolation}
\end{figure}

Figure~\ref{fig:isolation} shows how the density profile depends on
particular choice of the isolation criterion. In this case we selected few
hundred halos in the simulation Box80G with masses $M\approx
10^{12}\Msunh$. Qualitatively the same results are found for halos
with different masses. Conclusions are clear: More strict isolation
conditions result in smaller density in the outer parts of halos with
almost no effect inside the virial radius. Even outside of the virial
radius the difference are not that large. The difference between
distinct (not isolated) halos and halos, which have no massive
companions inside 2\Rvir, are not more than a factor $\sim 1.5$. To
large degree, this is not surprising because our isolated halos are
typical, i.e., more that 1/2 of the halos in this mass range are ``isolated''.

\begin{figure}[tb!]
\epsscale{1.0}
\plotone{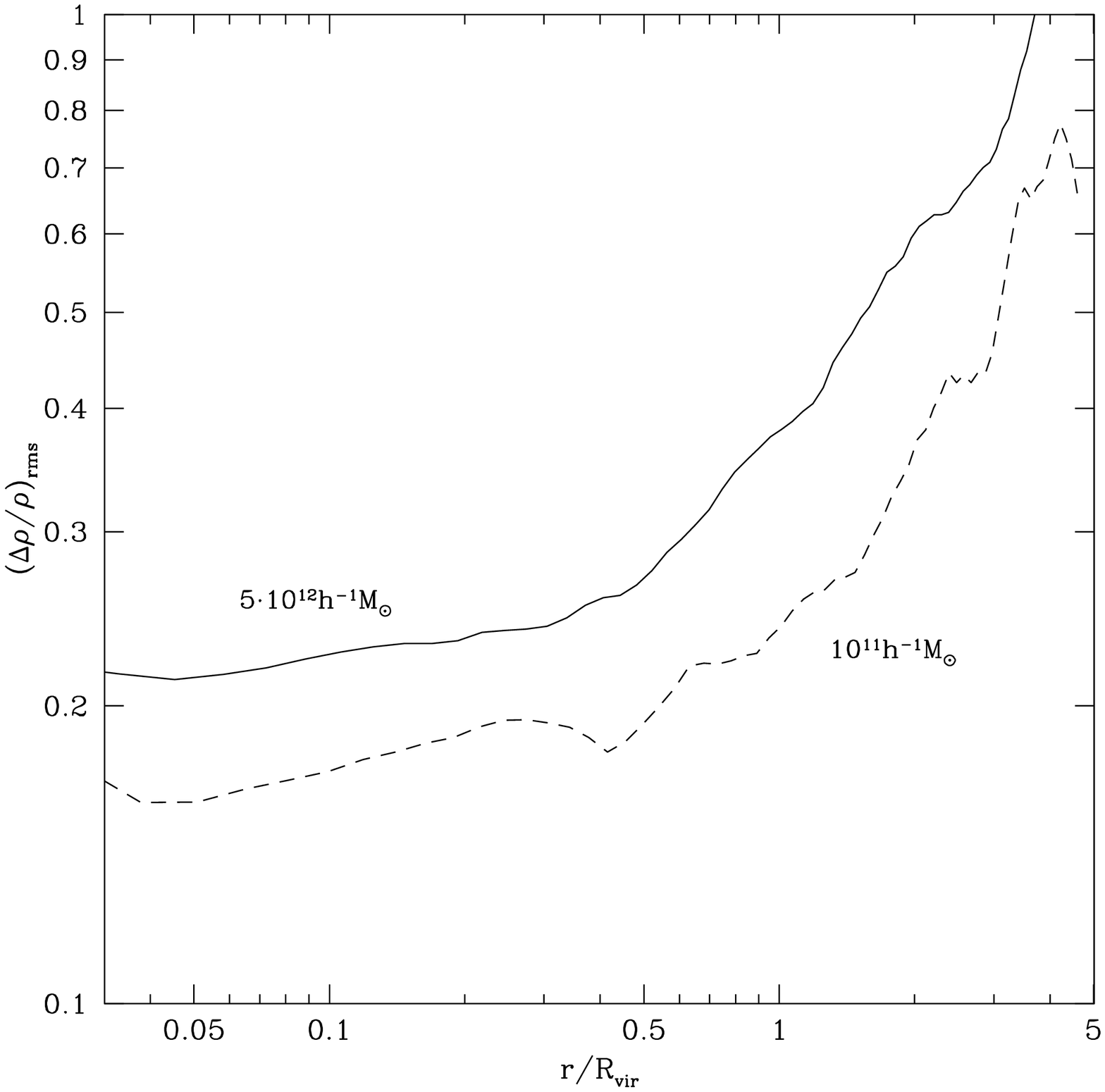}
\caption{RMS deviations of halo profiles from average profile for
halos of given mass. The curves in the plot show results for halos
with masses indicated in the plot. The deviations inside the formal virial
radius are clearly smaller than for the outer regions. Inside the virial
radius the deviations are smaller for halos with smaller mass.}
\label{fig:deviations}
\end{figure}

One of the misconceptions, which we had before starting the analysis of the
outer regions of dark matter halos is that at large distances the deviations from
halo to halo are so large that it is very difficult to talk about
average profile or a profile altogether. This appears to be not true.
Figure~\ref{fig:deviations} shows the halo to halo rms
deviations from the average density profile for halos of a given mass. In 
order to construct the
plot, we split the halo population into three ranges of concentrations and
found the average and deviations for each concentration bin. Then the
results of different concentrations were averaged. This splitting into
concentrations is needed only for the central region $r< 0.1\Rvir$
because here the average profile depends on the concentration. This
plot demonstrates that there is no drastic change in the deviations
at the virial radius. The deviations increase with the distance, but
they are not unreasonable.

\begin{figure}[tb!]
\epsscale{1.25}
\plotone{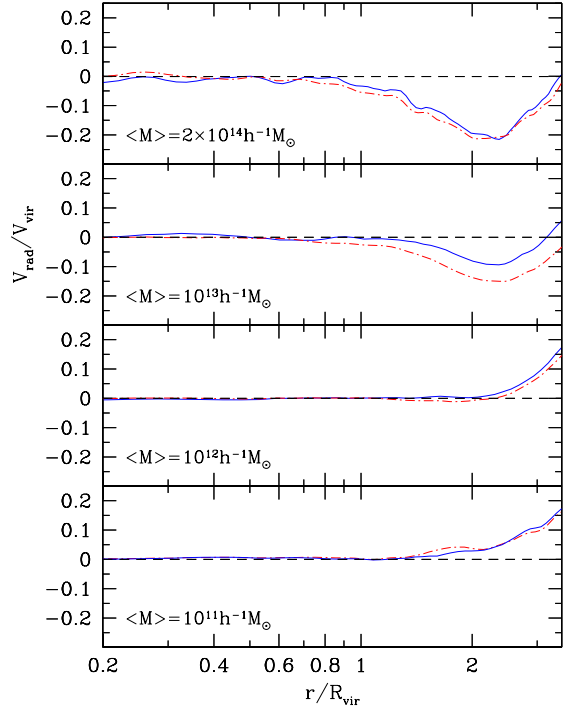}
\caption{Average radial velocities for halos with different virial
masses. The full curves show results for isolated halos and the
dot-dashed curves are for distinct halos.  The velocities are
practically zero within $2-3~\Rvir$ for halos with mass smaller than
$10^{12}\Msunh$.  The situation is different for group- and cluster-
sized halos (two top panels). For these large halos there are large
infall velocities, which amplitude increases with halo mass.}
\label{fig:infall}
\end{figure}

Figure~\ref{fig:infall} shows the average radial velocity profile for halos of
vastly different masses. We used many dozens of halos for each mass
range.  Just as in the case of the two individual halos in Figure~1
and 2, there is no systematic infall of material beyond formal virial
radius for small virial masses. The situation is different for group-
and cluster- sized halos (two top panels). For these large halos there
are large infall velocities, which amplitude increases with halo mass.

\section{Predictions from the spherical collapse model}
\label{sec:sph}

Here we obtain the predictions of the spherical collapse model 
\citep{GG72}
for the $\it{mean}$ halo density profile and compare 
them with the results found in the numerical simulations. The
spherical collapse model provides a relationship between the present 
nonlinear enclosed  density contrast, 
$\delta= \langle \rho(r)\rangle /\langle \rho_m\rangle -1$, within a
sphere of given radius $r$ and the 
enclosed linear density contrast, $\delta_l$, (i.e., the initial fluctuation extrapolated
to the present using the linear theory) within the same sphere. This
relationship along with the statistics of the initial
fluctuations furnishes definite predictions for the mean density profile.

We expect the spherical collapse model 
to give the $\it{mean}$ density
profile around virialized halos for sufficiently large radii -
significantly larger than $\Rvir$. 
In fact, for $r/\Rvir > 2.5$ we find
a good agreement between the prediction of the spherical model and the
$\it{mean}$ density profiles obtained from numerical simulations for
several mass ranges.
By $\it{mean}$ density profile we imply any representative profile
for a given mass range. We consider three different representative profiles:
the most probable profile, the mean profile and, what we call the
typical profile. This last profile is simply the mean profile in the
initial conditions spherically evolved.

Our procedure  to obtain the typical density profile
will be presented in full detail in Betancort-Rijo et
al. (2005). 
In essence, we use the statistics of the initial Gaussian field to
obtain the mean profile of the enclosed linear density contrast,
$\delta_{l}(q)$, around a proto-halo as a function of the Lagrangian
distance to the center, $q$. We then use the spherical collapse model
to obtain the present density contrast, $\delta(r)$, as a function of
the present comoving (i.e., Eulerian) distance to the center of the halo,
$r$.

If $Q$ is the virial radius in Lagrangian coordinates, then the 
enclosed linear contrast $\delta_{l}(q,Q)$ must satisfy the condition
that
$\delta_{l}(q=Q)=\delta_{vir},$
where $\delta_{vir}$ is the  linear density contrast
within the virial radius $\Rvir$ at 
the moment of virialization. The
present average enclosed fractional density $\delta(r=\Rvir)$ is equal to 
340 (the value of overdensity $\Delta_{\rm vir}$ used in our simulations
to define the virial radius). 
 To obtain $\delta_{vir}$ we must find the 
$\delta_{l}$ corresponding to a
present density contrast equal to 340. This condition leads to a 
$\delta_{vir}$ value of 1.9. We shall later comment on this
once we introduce the spherical collapse model. 

This condition
simply ensures that the proto-halo evolves into an object which  at
present is virialized within the prescribed virial radius $\Rvir$. If
this were the only constraint on the linear profile (equal to the
initial profile except for an overall factor), it have been shown
(see Section 4 in Patiri et al. 2004) that:

\begin{equation}
{\delta}_{l}(q,Q) \simeq \delta_{vir} 
\exp{\big[-b\big(\big(\frac{q}{Q}\big)^{2}-1\big)\big]} \equiv \delta_{0}(q,Q),
\label{eq:eq3}
\end{equation}

\noindent
where $b$ is a coefficient depending on $\Rvir$. It is given below.

However, although this
profile does not lead to wrong results, in order to achieve the accuracy 
required here and to obtain the correct dependence of the shape on mass
of the halos we must include an additional constraint. This constraint is:

\begin{equation}
{\delta}_{l}(q,Q) < \delta_{vir}  \quad   \forall~ q > Q.       
\label{eq:eq4}
\end{equation}
It means that there are no radii larger than $\Rvir$, where the
 density contrast is 340 or larger. Otherwise, the virial radius
would be larger than $\Rvir$. In the large mass limit
this condition becomes irrelevant, and 
the linear density contrast is given
by eq.(\ref{eq:eq3}). However, in the general case the mean linear profile,
${\delta}_l(q,Q)$, is somewhat steeper than $\delta_{0}(q,Q)$:

\begin{equation}
\delta _{l}(q,Q) = \delta_{0}(q,Q) -\frac{\sigma'(q,Q)\exp[-x^2]}{1-\frac{1}{2}{\rm erfc}[-x]}, 
\label{eq:eq5}
\end{equation}
\begin{equation}
   x \equiv \frac{\delta_{vir}-\delta_{0}(q,Q)}{\sqrt{2}\sigma'(q,Q)}
\label{eq:eq6}
\end{equation}

\noindent
with $\delta_0(q,Q)$ given in eq.(\ref{eq:eq3}), and

\begin{equation}
\sigma'(q,Q)\equiv \sqrt{\sigma^2(q)-\Big(\frac{\delta_l(q,Q)}
{\delta_{vir}}\Big)^2\sigma^2(Q)},
\label{eq:eq7}
\end{equation}

\noindent
where $\sigma(q)$ and $\sigma(Q)$ are the $rms$ linear density fluctuations in spheres with 
lagrangian radii $q$ and $Q$. We use $\sigma_8=0.9$ as in the simulations.

With these definitions we have for the previously defined function $b(\Rvir)$:

\begin{equation}
b(R_{\rm vir})=-\frac{1}{4}\frac{d{\rm ln}\sigma(Q)}{d{\rm ln}Q}\bigg|_{Q=R_{\rm vir}(340)^{1/3}}
\end{equation}

Using the linear density contrast eq.(\ref{eq:eq5}-\ref{eq:eq6}) we can now  obtain the nonlinear
density contrast, $\delta(r)$. For any given radius $r$ the nonlinear density contrast $\delta(r)$ is given
by the solution to the equation:

\begin{equation}
\delta_{l}(\delta(r))=\delta_{l}(q,Q),
\label{eq:eq8}
\end{equation}
\begin{equation}
q \equiv r(1+\delta(r))^{1/3} \quad, \quad Q \equiv R_{vir}(340)^{1/3}.
\label{eq:eq9}
\end{equation}

\noindent
The derivation of this eq.(\ref{eq:eq8}) may be found in
Patiri et al.(2004), although here the equation is presented in a
slightly different form. The left hand side of this equation is 
simply the relationship, $\delta_{l}(\delta)$, between the present and 
the linear $\delta$ value in the spherical collapse model 
(see Sheth \& Tormen 2002) evaluated at $\delta=\delta(r)$. For
the cosmology considered here we have for $\delta_{l}(\delta)$:

\begin{eqnarray}
   \delta_{l} (\delta) &=& \frac{1.676}{1.68647} \left[ 1.68647
   - \frac{1.35}{(1+\delta)^{2/3}} \right.
   \nonumber \\
   &-& \frac{1.12431}{(1+\delta)^{1/2}}
   + \left. \frac{0.78785}{(1+\delta)^{0.58661}} \right].     \label{eq:ST}
\end{eqnarray}
The right hand side of eq.(\ref{eq:eq8}), which
depends on $\delta(r)$ through $q$, is given by expression (\ref{eq:eq3}). 


Inserting eq.(\ref{eq:eq5}-\ref{eq:eq6}) into 
eq.(\ref{eq:eq8}) with $\delta_{vir}=1.9$,
using for $b$ the values 0.186 and 0.254 for the two masses $6.5
\times 10^{10} \Msunh$ and $3 \times 10^{12} \Msunh$ discussed here,
and solving for $\delta(r)$, we obtain the profiles for the present
 enclosed  density contrast, $\delta(r)$. To
obtain the density $\delta'(r)$ at a given radius $r$
 we simply need to take a derivative:

\begin{equation}
(1+\delta'(r)) \equiv \frac{\rho(r)}{\langle \rho_m \rangle } \quad,\quad 
 \delta'(r)=\frac{1}{3}\frac{1}{r^{2}}\frac{d}{dr}r^{3}\delta(r),
\label{eq:eq10}
\end{equation}

\noindent
where  $\langle \rho_m \rangle$ is the
mean matter density of the Universe.

We must now comment on the value of $\delta_{vir}$ that we use. If the standard spherical
collapse model were valid down to the virial radius we could use expression (\ref{eq:ST}) to
obtain it:

\begin{displaymath}
\delta_{vir}=\delta_{l}(340)=1.614
\end{displaymath}
However, we know that this is not true because bellow $\sim 2$ virial radius there is substantial
amount of shell-crossing that render the mentioned model unappropiate. This causes $\delta$ to grow
much more slowly with $\delta_{l}$, so that when $\delta$ takes the value 340, $\delta_{l}$ takes
the value 1.9 \citep{Beta04}. 

To obtain the predictions for the most probable and mean profiles, the
probability distribution, $P(\delta, s)$, for the value of $\delta$ at
a given dimensionless radius $s\equiv r/R_{vir}$ is needed.
This distribution is derived in Betancort-Rijo
et al.(2005) along the same lines we have described for the the
typical profile. Here we simply use it to compute the most probable,
$\delta(s)_{\rm prob}$ profile which is is simply given by the value
at which $P(\delta, s)$ has its maxima. The mean profile,
$<\delta(s)>$ is obtained from:

\begin{equation}
<\delta(s)>=\int_{-1}^{\infty} P(\delta,s) d\delta.
\end{equation}

\begin{figure*}[tb!]
\epsscale{2.0}
\plotone{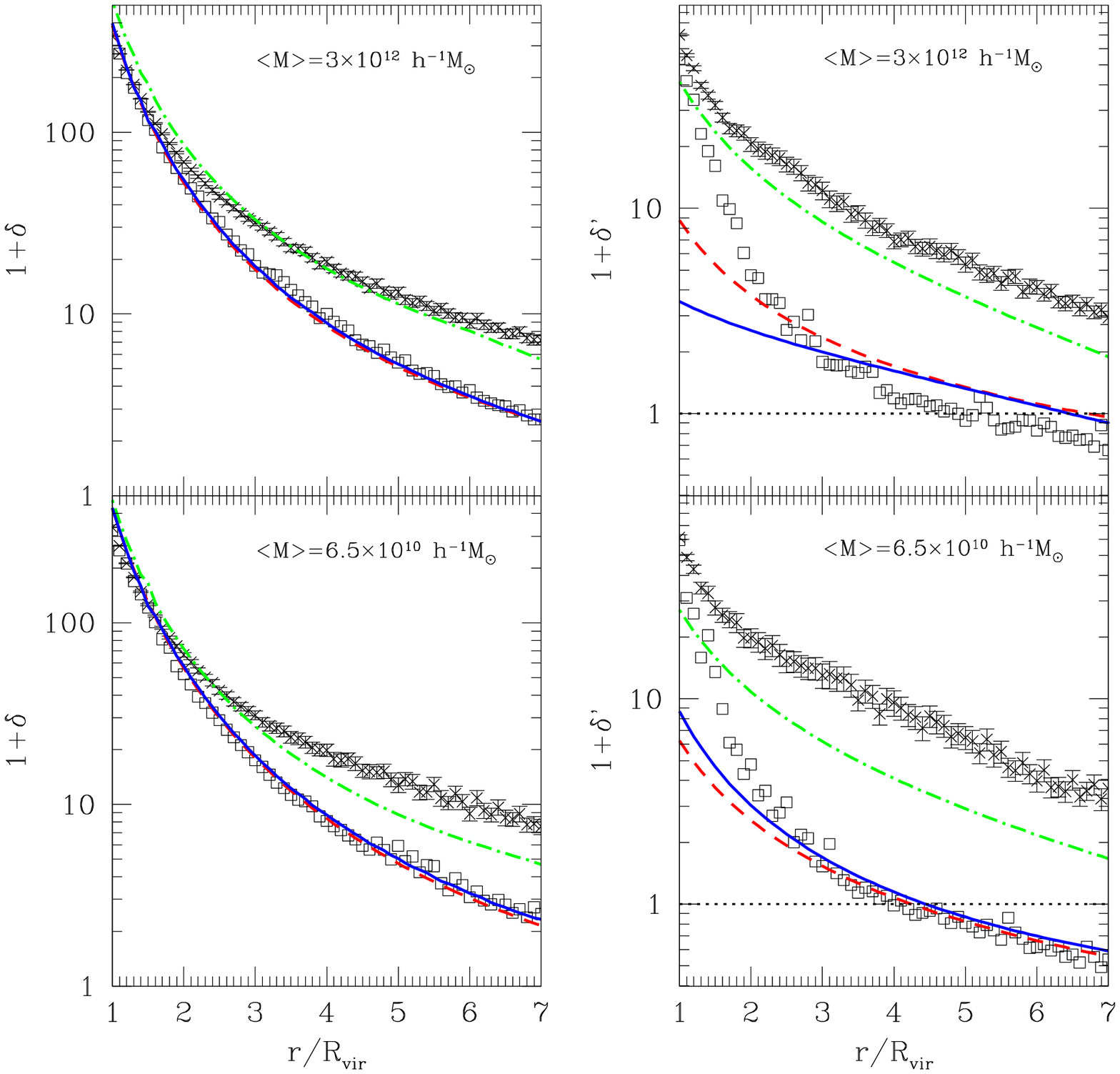}
\caption{The most probable (squares) and the mean (crosses) halo density profiles 
up to 7$\Rvir$ for the two masses $(<M>=3\times10^{12} \Msunh$ and 
$<M>=6.5\times10^{10} \Msunh$ in our simulations. We compare 
the simulated data with the predictions from the spherical collapse model for the
most probable (solid line) and typical profiles (dashed line) which are
almost indistinguishable and the mean profiles (dot-short dashed line). We have estimated that the errors in the most probable 
values of $\delta$ and $\delta'$ are about $25\%$ larger than the error showed
in the plots for their mean values.}
\label{fig:fig6}
\end{figure*}

In practice, since the approximation we use for $P(\delta, s)$ has
an unduly long tail, we have to truncate it artificially. So, we only
integrate up to a $\delta$ value where $P(\delta, s)$ has fallen to a
twenty fifth of its maximum value. The results are presented in 
Figure~\ref{fig:fig6}, where we have
computed, both for $\delta$ and for $\delta'$, the most probable
(squares) and mean (crosses) density profiles found in our simulations
for two mass intervals with the mean values equal to the masses used
in the theoretical derivation.  We have taken $277$ halos in the mass
range $(6.5\pm1.5)\times10^{10}\Msunh$ from $Box80S$ and $654$ halos
in the mass range $(3\pm1)\times10^{12}\Msunh$ from $Box80G$. No
isolation criteria was used. In Table~\ref{tab:tab3} we list for the
mean halo with mass $<M>=3\times10^{12}\Msunh$ the estimations of the
most probable and mean value of the density at different radii compare
with that from the spherical collapse model for the most probable, the
mean and the typical profiles.

\begin{figure}[tb!]
\epsscale{1.0}
\plotone{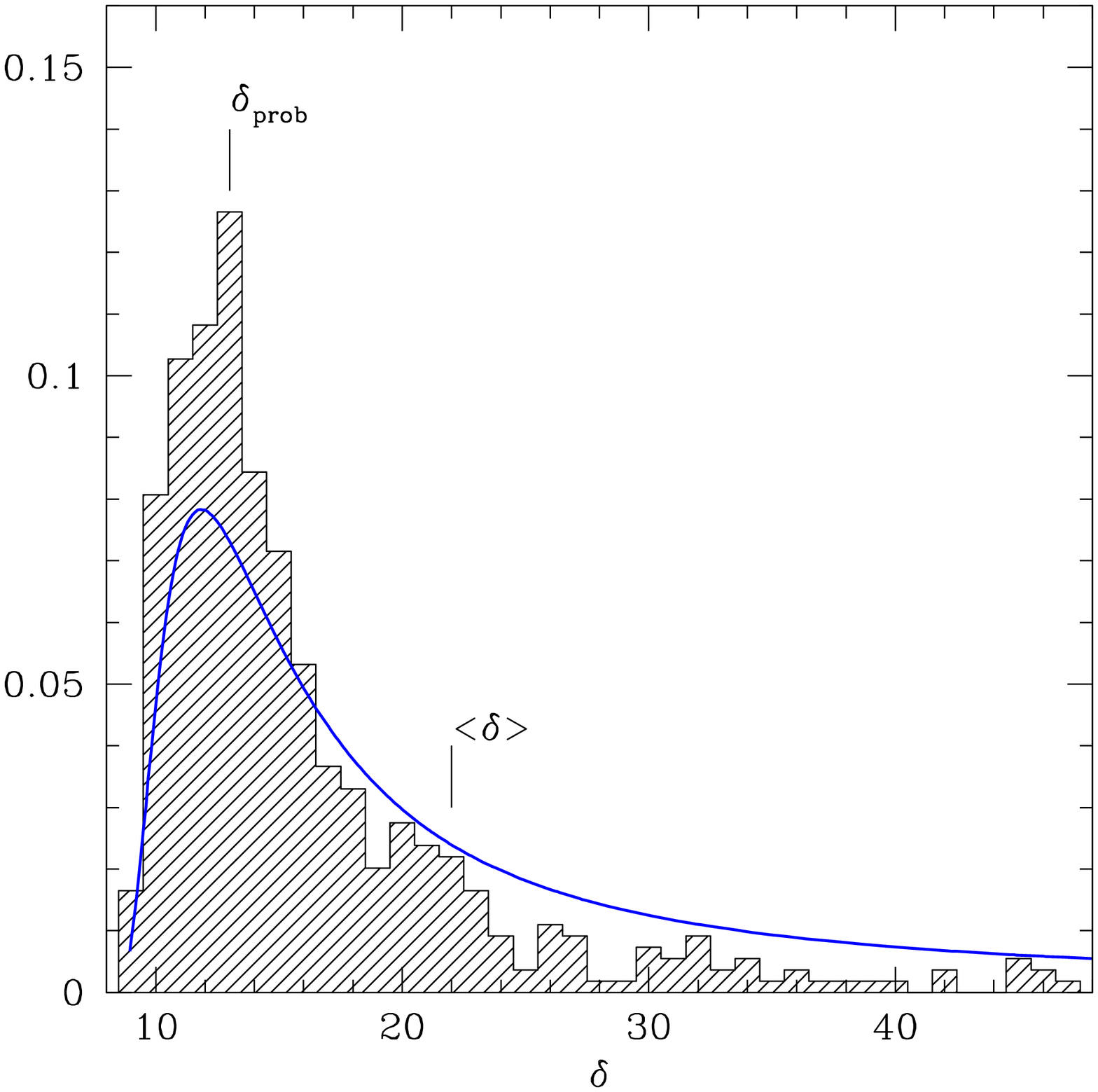}
\caption{Distribution of the fractional cumulative density $\delta$ inside 
$3.5\pm0.05 \Rvir$ for the mean halo of mass $<M>=3\times10^{12} \Msunh$. We show 
for comparison the theoretical prediction of $P(\delta, s)$ as well as we give 
the most probable $\delta_{\rm max}$ and mean value $<\delta>$ of the distribution.
We display the density distribution from its minimum value up to 1$\sigma$ from 
its mean $<\delta>$ (8\% of the values of the distribution are beyond 1$\sigma$).}
\label{fig:fig7}
\end{figure}

In Figure~\ref{fig:fig6} one can see that beyond 2 virial radius the mean and 
the most probable profiles, both for $\delta$ and for $\delta'$, differ 
considerably. This is due to the fact that for this radii the probability 
distribution for $\delta$, $P(\delta,s)$, is rather wide, with a long upper tail. 
This can be seen in Figure~\ref{fig:fig7} were this distribution is shown inside 
$3.5\pm0.05$ virial radius for the mass $<M>=3 \times 10^{12}\Msunh$. We show 
for comparison the theoretical prediction for $P(\delta, s)$ as well as we give 
the most probable $\delta_{\rm max}$ and mean value $<\delta>$ of the distribution.

It is apparent from Figure~\ref{fig:fig6} that the $\delta'$ profiles
are steeper for smaller masses, so that they go below the background
at smaller $r/R_{vir}$ and reach larger underdensities. The $\delta$
profiles are also steeper for smaller masses although the difference
is, obviously, much smaller. We have found that the theoretical
prediction for the typical and the most probable profile are, in
general, almost indistinguishable (see Table~\ref{tab:tab3}). They are
both found to be in very good agreement with the most probable
$\delta$ profile found in the numerical simulations beyond two virial
radii. There is also qualitative agreement between the predictions of
the most probable $\delta'$ profiles and those found in the numerical
simulations. It must be noted, however, that by predicted $\delta'$
profile we understand simply the one obtained from the corresponding
$\delta$ profile by means of relationship given in eq.(\ref{eq:eq10}). Note that
this is not the same as the most probable profile for $\delta'$ (see
Figure~\ref{fig:fig6}). This is due to the fact that the most probable
$\delta'$ value at a given $s (\equiv r/R_{vir})$ corresponds to a
different halo than the most probable $\delta$ value at the same
$s$. This explains that, while the prediction for $\delta_{prob}$
agrees very well with the simulations, the agreement is not so good
for $\delta'_{prob}$. The $\delta'_{prob}$ obtained from
$\delta_{prob}$ by means of expression eq.(\ref{eq:eq10}) is not a proper prediction
but an indicative value, since we can not envisage a feasible
procedure to obtain a proper prediction. On the contrary, the mean
profiles $<\delta>$ are exactly related to $<\delta'>$ by means of
expression eq.(\ref{eq:eq10}). The predictions for both profiles are in good
agreement with the numerical simulations, showing a much flatter
profile beyond 2 virial radius than those corresponding to
$\delta_{\rm prob}$ and $\delta'_{\rm prob}$.  This agreement is remarkable given
the fact that the expression used for $P(\delta,s)$ is only a first
approximation (see Betancort-Rijo et al. 2005).

\begin{table*}[tb]
\tablenum{3}
\label{tab:tab3}
\caption{Comparison between the simulated $\it{mean}$ halo density profile and the 
theoretical predictions from the spherical collapse model for the mass 
$<M>=3\times10^{12} \Msunh$.}
\begin{center}
\small
\begin{tabular}{lccccccccccccl}
\tableline\tableline\\
\noalign{Numerical Simulations \hskip 10em Spherical Collapse}
\multicolumn{1}{c}{$r/R_{\rm vir}$} & 
\multicolumn{1}{c}{}        &
\multicolumn{1}{c}{$<\delta>$} &
\multicolumn{1}{c}{$\delta_{\rm prob}$} &
\multicolumn{1}{c}{$<\delta'>$} &
\multicolumn{1}{c}{$\delta'_{\rm prob}$} &
\multicolumn{1}{c}{}        &
\multicolumn{1}{c}{}        &
\multicolumn{1}{c}{$<\delta>$} &
\multicolumn{1}{c}{$\delta_{\rm prob}$} &
\multicolumn{1}{c}{$\delta_{\rm t}$} &
\multicolumn{1}{c}{$<\delta'>$} &
\multicolumn{1}{c}{$\delta'_{\rm prob}$} &
\multicolumn{1}{c}{$\delta'_{\rm t}$}  
\\
\tableline
\\
1.0 &  & 337  & 323  & 69.1 &  52.4   &  &  & 531  &  398  & 385   & 40.4 & 2.5    &  7.8\\ 
1.5 &  & 129  & 116  & 30.8 &  15.1   &  &  & 184  &  115  & 118   & 22.6 &  1.9    &  4.3\\ 
2.0 &  & 67.7 & 54.6 & 19.5 &  3.73   &  &  & 84.8 &  53.7 & 51.7  & 14.7 & 1.5    &  2.8\\ 
2.5 &  & 43.2 & 31.1 & 15.4 &  1.6    &  &  & 49.3 &  28.6 & 27.6  & 10.3 & 1.2    &   1.9\\ 
3.0 &  & 30.8 & 17.4 & 11.2 &  0.78   &  &  & 32.2 &  17.3 & 16.6  & 7.6 & 1.0   &  1.4\\ 
3.5 &  & 22.1 & 12.6 & 8.5  &  0.58   &  &  & 22.6 &  11.3 & 10.9  & 5.8 & 0.79   &  0.97\\ 
4.0 &  & 18.4 & 9.0  & 5.9  &  0.18   &  &  & 16.8 &  7.9  & 7.6  & 4.5 & 0.61   &  0.70 \\ 
4.5 &  & 14.1 & 5.8  & 5.5  &  0.09   &  &  & 13.0 &  5.8 & 5.5  & 3.5 & 0.46   &  0.50 \\ 
5.0 &  & 12.2 & 4.4  & 4.3  &  -0.08  &  &  & 10.4 &  4.3 & 4.1  & 2.7 & 0.32   &  0.34 \\ 
5.5 &  & 9.4  & 3.5  & 3.3  &  -0.16  &  &  & 8.5  &  3.3 & 3.2  & 2.1 & 0.20   &  0.22 \\ 
6.0 &  & 7.9  & 2.8  & 3.1  &  -0.17  &  &  & 7.1  &  2.5 & 2.5  & 1.6 & 0.09   &  0.12 \\ 
6.5 &  & 6.8  & 2.1  & 2.5  &  -0.22  &  &  & 5.7  &  2.0 & 2.0  & 1.2 & -0.008 &  0.03 \\ 
7.0 &  & 6.2  & 1.6  & 1.9  &  -0.33  &  &  & 4.6  &  1.6 & 1.6  & 0.9 &  -0.10 &  -0.04 \\ 
\\             
\end{tabular}
\tablecomments{
The symbols $<\delta>$, $\delta_{prob}$, $\delta_{t}$ stand,
 respectively, for mean, most probable, and typical averaged enclosed fractional overdensity.
 The corresponding primed simbols are for the local fractional overdensities at given radius.}
\end{center}
\end{table*}

It is interesting to note, as we have previously pointed out, that
larger masses have somewhat shallower profiles. In order to
predict this trend correctly we must use the initial profile given by
eq.(\ref{eq:eq5}-\ref{eq:eq6}). If we dropped the second constraint 
(that is given in eq.(\ref{eq:eq4}) and use in eq.(\ref{eq:eq8}) the initial 
profile given by eq.(\ref{eq:eq3}), which
corresponds to high mass objects, the prediction would be the
opposite. The reason for this being that, in this limit, the initial
profile depend on mass only through $c$ which increases with
increasing mass, thereby leading to steeper profiles for larger
masses.

As we stated in the introduction, the computations of the $\it mean$
profiles (in fact, the typical profile) around halos has independently
been made by \citet[][]{Barkana}. He used the spherical model and, in
principle, imposed on the initial profile the same constraint as we
do. The computing procedure he followed was somewhat different
involving some approximations. He do not give explicitly the equation
defining the profile, so that accurate comparison with our results are
not possible. Furthermore he used different values of $\delta_{vir}$,
$\Delta_{vir}$. However, his results are in good qualitative agreement
with our predictions for the typical profile.

\begin{figure}[tb!]
\epsscale{1.0}
\plotone{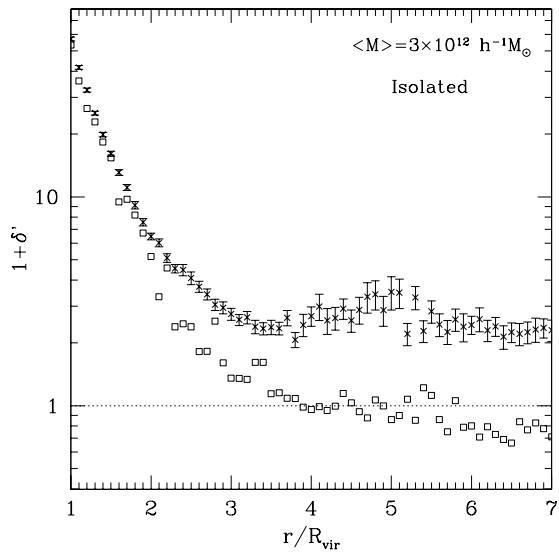}
\caption{The mean (crosses) and most probable values (squares) of the
local density profile for the mean isolated
dark matter halo with mass $<M>=3\times10^{12}\Msunh$.}
\label{fig:fig8}
\end{figure}

We have so far considered randomly chosen halos, i.e. without
isolation criteria.  When halos are chosen according with an isolation
criteria they differ from the randomly chosen ones in two respects. On
the one hand, the probability distribution for $\delta$ at a given
value of $s$ is narrower, so that most profiles cluster around the
most probable one. Therefore, the difference between the mean and the
most probable density profile becomes smaller. On the other hand,
isolated profiles lay, on average, on somewhat more underdense
environment than non-isolated ones, so that their most probable
profiles are slightly steeper. Both effects may be seen by comparing
Table~\ref{tab:tab3} and Table~\ref{tab:tab4} or the upper right
pannel in Figure~\ref{fig:fig6} with Figure~\ref{fig:fig8} where we
show the local density profile for the isolated mean halo density
profile for the mass range $3\pm1\times10^{12} \Msunh$. In this mass
range we have selected halos that do not have a companion with mass
larger than 10\% of the halo mass within $4\Rvir$. In total there are
$156$ halos, i.e. one quarter of all halos in this mass range.

\begin{table}[tb]
\tablenum{4}
\label{tab:tab4}
\caption{The isolated $\it{mean}$ halo density profile for the mass 
$<M>=3\times10^{12} \Msunh$. The symbols are the same as table 3.}
\begin{center}
\small
\begin{tabular}{lccccccccccccl}
\tableline\tableline\\
\multicolumn{1}{c}{$r/R_{\rm vir}$} & 
\multicolumn{1}{c}{}        &
\multicolumn{1}{c}{$<\delta>$} &
\multicolumn{1}{c}{$\delta_{\rm prob}$} &
\multicolumn{1}{c}{$<\delta'>$} &
\multicolumn{1}{c}{$\delta'_{\rm prob}$}  
\\
\tableline
\\
1.0 &  & 336  & 323  & 56.1 & 52.4    \\ 
1.5 &  & 119  & 119  & 15.1 & 14.4    \\ 
2.0 &  & 55.4 & 65.6 & 5.5 &  4.2    \\ 
2.5 &  & 30.1 & 29.4 & 3.1 &  1.4    \\ 
3.0 &  & 18.5 & 18.2 & 1.8 &  0.36   \\ 
3.5 &  & 12.3 & 11.0 & 1.4 &  0.14   \\ 
4.0 &  & 8.5 & 7.0 & 1.7 & -0.04  \\ 
4.5 &  & 6.4 & 6.0 & 1.6 &  0.03 \\ 
5.0 &  & 5.2 & 4.6 & 2.5 & -0.14  \\ 
5.5 &  & 4.5 & 2.9 & 1.8 &  0.12  \\ 
6.0 &  & 3.9 & 2.7 & 1.4 & -0.20 \\ 
6.5 &  & 3.4 & 2.0 & 1.3 & -0.34  \\ 
7.0 &  & 3.0 & 1.5 & 1.3 & -0.30  \\ 
\\             
\tableline
\end{tabular}
\end{center}
\end{table}

\section{Conclusions and Discussions}
\label{sec:fin}

We perform a detailed study of the  density profiles of 
isolated galaxy-size dark matter halos in high resolution cosmological 
simulations. We devote careful consideration mainly to the halo outer
structure beyond the formal virial radius $\Rvir$. We find that the
3D S\'ersic three-parameter approximation provides excellent good density fits
for these dark matter halos up to 2-3$\Rvir$. These profiles do not display
an abrupt change of shape beyond the virial radius. The halo-to-halo rms deviations
from the average profile for halos of a given mass show that there is no a drastic
change in the deviations at the virial radius. We show that these density
profiles differ considerably from the NFW density profile beyond 3$\Rvir$ where the 
density profile are slower than $r^{-2}$.  This result must not be seen as
a contradiction when is compared with the $r^{-3}$ NFW fall-off at large radii
since we must remember that the NFW analytical formula was proposed and extensively
tested to describe the structure of virialized dark matter halos within $\Rvir$. 
Although surprising, it is customary, for example, to see that the weak lensing analysis 
is often done with the NFW fit extrapolated to distances well beyond $\Rvir$,
up to large distances of several virial radii (up to few Mpc). This approach may not 
be accurate enough given the results presented in this work. 

We also find that the isolated galaxy-size halos display all the
properties of relaxed objects up to 2-3$\Rvir$. In addition to their
relatively smooth density profiles seen at large radii, by studying
halos average radial velocities, we find that there is no indication
of systematic infall of material beyond the formal virial radius. The
dark matter halos in this mass range do not grow as one naively may
expect through a steady accretion of satellites, i.e., on average
there is no mass infall.  This is strikingly different for more
massive halos, such as group- and cluster-sized halos which exhibit
large infall velocities outside of the formal virial radius. For large
halos the amplitude of the infall velocities increases with halo mass.

For larger radii beyond 2-3 formal virial radius we combine the
statistics of the initial fluctuations with the spherical collapse
model to obtain predictions of the mean halo density profiles for
halos with different masses. We consider two possibilities: the most
probable and the mean density profiles. We find that the most probable
profile obtained from our simulations is in excellent agreement with
the predictions from the spherical collapse model beyond 2-3 virial
radius. For the mean density profile the predictions are not so
accurate. This is due to the fact that the approximation, which we are
using for the distribution of $\delta$ at a given radius, has an
artificially long tail (we are presently working on a better
approximation). Even so, the predictions are qualitatively good and
quantitatively quite acceptable. We think that the discrepancies
between the data and the predictions at radii smaller than 2-3 virial
radii are due to the fact that these inner shells are affected by the
shell-crossing. We hope that an appropiate treatment of this
circunstance will lead to accurate predictions for all radii, so that the
mean spherically averaged profiles may be understood in terms of the
spherical collapse. We find the results presented here very
encouraging in this respect and we are currently working along this line.

\acknowledgements We thank Yehuda Hoffman and Simon White for
stimulating discussions.  F.P. is a Ram\'on y Cajal Fellow at the IAA
(CISC). F.P. want to thank the hospitality and financial support of
the Instituto de Astrofisica de Canarias and the New Mexico State
University where part of this work was done. F.P., S.P. and S.G. would
like to thank Acciones Integradas for supporting the German-Spanish
collaboration. S.G. acknowledges support by DAAD. A.K. acknowledges
support of NASA and NSF grants to NMSU.  Computer simulations have
been done at the LRZ Munich, NIC J\"ulich and the NASA Ames. Part of the data
analysis have been carry out at CESGA.

\newpage

\end{document}